\documentclass[onecollarge]{svjour2}       
\usepackage{graphicx}
\usepackage{amssymb}
\usepackage{color}
\begin{document}

\title{Ground state and excitation properties of soft-core bosons}

\author{Tommaso Macr\`i \and Sebastiano Saccani \and Fabio Cinti}


\institute{Tommaso Macr\`i \at
		   Max Planck Institute for the Physics of Complex Systems, N\"othnitzer Stra\ss e 38, 01187 Dresden, Germany\\ 
              Present address:
              Max-Planck-Institut f\"{u}r Quantenoptik, Hans-Kopfermann-Str.~1, 85748 Garching, Germany, and
		   QSTAR, Largo Enrico Fermi 2, 50125 Firenze, Italy\\
              \email{macri@cqstar.eu}         
         \and         
             Sebastiano Saccani \at
             SISSA Scuola Internazionale Superiore di Studi Avanzati and DEMOCRITOS National Simulation Center,
             Istituto Officina dei Materiali del CNR Via Bonomea 265, I-34136, Trieste, Italy\\
             \email{sebastiano.saccani@gmail.com}         
         \and
             Fabio Cinti \at 
             National Institute for Theoretical Physics (NITheP), Stellenbosch 7600, South Africa \\
             \email{cinti@sun.ac.za}                 
}

\date{Received: date / Accepted: date}

\maketitle

\begin{abstract}
We study the physics of soft-core bosons at zero temperature in 
two dimensions for a class of potentials that could be realised in 
experiments with Rydberg dressed Bose-Einstein condensates. 
We analyze the ground state properties of the system in detail and 
provide a complete description of the excitation spectra in both superfluid, supersolid and crystalline phase
for a wide range of interaction strengths and densities. 
In addition we describe a method to extract the transverse gapless 
excitation modes in the phases with broken translational symmetry within the framework of
path integral Monte Carlo methods.




\end{abstract}

\section{Introduction}
\label{intro}



A considerable number of theoretical studies have been lately devoted to  
investigate ultra-cold gases characterised by non-local interactions \cite{Bloch2008},
such as dipolar bosons and fermions \cite{Lahaye2009}, 
Rydberg atoms \cite{Saffman2010}, and polar molecules \cite{Micheli2007}.  
From the experimental perspective this field of research is increasingly turning into  
a perfect playground for realizing exotic phases with non-trivial broken symmetries 
\cite{baranov_2012,pupillo2008,Schausz:2012fk}, 
which are hardly observable in typical condensed matter systems.
A primary example is the so called supersolid phase.
Over the past few decades the existence of supersolidity was subjected to an extensive 
experimental and theoretical work, mostly dominated by studies on solid $^4$He 
\cite{RevModPhys.84.759,Andreev1969,Chester1970,PhysRevLett.25.1543,Meisel1992,Kim04a,Kim04b,PhysRevLett.109.155301}. 
From a theoretical viewpoint supersolidity can be understood as a phase which  
simultaneously breaks both translational symmetry (crystalline order) and 
global gauge symmetry that enables long-range phase coherence and thereby superfluidity of the system \cite{RevModPhys.84.759}. 
Some recent theoretical investigations \cite{PhysRevLett.104.195302,PhysRevLett.104.223002} 
proposed that a supersolid phase may emerge from a Bose-Einstein condensate where 
particles are off-resonantly excited to a Rydberg state.
Under such conditions one can engineer effective two-body soft-core potentials which
at large distances decay with the usual dipolar or Van der Waals power law, 
while at shorter ones  approach a constant finite value --in contrast to the usual pure long range interactions--
as a consequence of the dipole blockade effect \cite{PhysRevLett.87.037901}.  
Recently, quantum Monte Carlo (QMC) calculations in continuum space 
provided a complete description of the ground state phase diagram for soft-core bosons,
confirming the presence of a cluster supersolid phase in two \cite{Cinti2010b} 
and three dimensions \cite{PhysRevA.88.033618}. 
Such superfluid clusters of a sufficient amount of particles per single site (high density regime) display physical properties
which are also efficiently described by mean field approaches based on the solution 
of an effective non local Gross-Pitaevskii equation 
\cite{Pomeau1994,PhysRevA.87.061602,PhysRevA.88.033618}. 

Lately, Cinti et al. \cite{cinti_nat_comm} investigated the ground state features of soft-core bosons at
lower densities where correlations can still give rise to supersolidity with a lower cluster occupancy.
Interestingly, contrarily to the predictions of the mean field theory, at low densities superfluidity 
increases linearly with the number of zero-point defects in the ground state. 
This picture is indeed fully consistent with the mechanism
proposed long ago by Andreev, Lifshitz \cite{Andreev1969} and Chester \cite{Chester1970} (ALC), where
superfluidity may, in fact, emerge through the formation of delocalized defects in the crystalline ground state. 

The presence of two different microscopic mechanisms underlying the same macroscopic effect 
in different regimes of the parameter space, opens up other intriguing questions regarding static and dynamical 
properties of a supersolid. Recently it was shown for instance that 
the excitation spectrum of the supersolid displays well defined Goldstone modes \cite{PhysRevA.87.061602,PhysRevLett.108.175301}, 
which emerge as a result of global gauge and translational symmetry breaking, respectively. 
It is not clear however how those modes show up and possibly deviate from the mean field approximation
in the low density limit where correlations in the ALC regime are expected to be significant. 

In the present paper we shall compare mean field results in the high and low density regime 
for a wide range of interaction strengths with exact QMC calculations for interaction potentials 
which can be simulated with Rydberg dressed atoms. 
Moreover we extend previous results concerning the spectrum of the elementary excitations 
extracting the transverse modes in the supersolid and crystalline phase with a method that 
exploits the crystalline arrangement of the ground state wavefunction. 
These calculations turn out to be essential to capture the range of 
validity of mean field calculations not only in the low density regime, but also at larger filling of the crystalline lattice
where optical modes systematically shifts upwards the energy of the longitudinal mode in a way that is
not controllable with advanced techniques of inversion of the intermediate scattering function \cite{Vitali2010}.



The paper is organized as follow: In the next Section, after having introduced the
system Hamiltonian, we discuss (Section~\ref{sec:2.1}) the mean-field
approach based the Gross-Pitaevskii equation with
a non-local Rydberg-dressed potential and define 
the Bogoliubov-de Gennes equations to study the excitation spectrum.
QMC techniques are exposed in Section~\ref{sec:2.2}, focusing the attention on 
the method we employed to calculate the transverse excitations in the solid-like phase. 
In Section~\ref{sec:3} we show the results regarding the ground state properties across 
the liquid-solid transition (Section~\ref{sec:3.1}), including a detailed analysis of the effects 
due to fluctuations in the uniform phase at high densities. In Section~\ref{sec:3.2}  
we show the excitation spectra in the limit of high and low densities. 
In Section~\ref{sec:4}  we draw the conclusions and examine future extensions of the present work.


%

\section{Physical system and Methodologies}
\label{sec:2}

We consider a system of bosons in two dimensions at zero temperature with  mass $m$ and positions ${\bf q}_i$, 
described by the Hamiltonian
\begin{equation}\label{MC}
\hat H = \sum_i  -\frac{\hbar^2}{2m}  \nabla^2_i +
\sum_{i<j} V({\bf q}_i-{\bf q}_j).
\end{equation}
The interaction is of soft-core nature and reads explicitly as 
$V({\bf r})=\frac{U}{r^6+R_c^6}$, being $U$ and $R_c$ strength and range of the interaction potential, respectively . 
For large inter-particle distances this
potential shows a Van Der Waals-like power law decay, whereas for vanishing separation
particles acquire a finite interaction energy $U/R_c^6$.
Upon scaling lengths by $R_c$ and energies by $\hbar^2/m R_c^2$, the zero temperature physics, 
determined by Eq.(\ref{MC}), depends only on two dimensionless parameters: an effective interaction 
strength $\alpha^{\prime}=U m /\left(\hbar^2 R_c^4\right)$ and the dimensionless density $n\, R_c^2$.

\subsection{Mean-Field approach}
\label{sec:2.1} 

Here we first review the mean field description 
\cite{PhysRevLett.104.195302,PhysRevLett.72.2426}
and the phases emerging from Eq.(\ref{MC}) at zero temperature \cite{cinti_nat_comm}.
In mean field theory the system dynamics is described by a non-local Gross-Pitaevskii equation (GPE), which 
reads in reduced units:
\begin{equation} 
\label{GPNH}
i \partial_t \psi({\bf r},t) = \left(-\frac{ \nabla^2}{2} + \alpha \int {\rm d}{\bf r}^{\prime} U({\bf r}-{\bf r}^{\prime}) 
|\psi({\bf r}^{\prime},t)|^2\right) \psi({\bf r},t) \;,
\end{equation}
where ${\bf r} = {\bf q}/R_c$, $U({\bf r})=\frac{1}{1+r^6}$, and $\alpha = \alpha' n\, R_c^2=m\,n\, U /\left(\hbar^2 R_c^2\right)$ 
is a dimensionless interaction strength that 
determines the ground state properties and the excitation dynamics.
The energy associated to the state described by the wave function 
$\psi({\bf r},t)=e^{-i\mu t}\psi_0({\bf r})$ in Eq.(\ref{GPNH}) can be derived from the GP energy functional:
\begin{equation}
\label{GPenergy}
H = \int {\rm d}{\bf r}\; \frac{1}{2} \left| \nabla \psi_0(\mathbf{r})\right|^2 + 
\frac{\alpha}{2} \int {\rm d}{\bf r} \, {\rm d}{\bf r}^{\prime}\; |\psi_0({\bf r})|^2 U({\bf r}-{\bf r}^{\prime}) 
|\psi_0({\bf r}^{\prime})|^2 \;.
\end{equation} 
In order to numerically determine the location of the transition from a uniform to a modulated ground state, we first expand the wavefunction 
$\psi_0(\mathbf{r})$ in Fourier series:
\begin{equation} \label{var}
\psi_0(\mathbf{r})=\sum_\mathbf{Q} C_\mathbf{\, Q}\; e^{i\, \mathbf{Q}\cdot \mathbf{r}}, 
\end{equation}
where $\mathbf{Q}=n\, \mathbf{b}_1+m\, \mathbf{b}_2$ with $n,m$ integers and $\mathbf{b}_1=\frac{2\pi}{a}\left(1,-\frac{1}{\sqrt{3}}\right)$,
$\mathbf{b}_2=\frac{2\pi}{a}\left(0,\frac{2}{\sqrt{3}}\right)$ are the reciprocal lattice basis vectors of a triangular lattice. 
We then substitute Eq.(\ref{var}) into Eq.(\ref{GPNH})
and iteratively solve the non-linear equations for $C_\mathbf{\, Q}$
until convergence is reached  \cite{PhysRevB.86.060510}. This procedure allows to determine the optimal lattice spacing,
the chemical potential and the coefficients $C_\mathbf{\, Q}$.

The validity of the above mean field theory is limited to the regime of high densities, that is,
where the depletion of the condensate remains small for a wide range of interaction strengths. 
Similar considerations were recently done on analogous soft-core step-like potentials leading 
to results in very good agreement with QMC simulations \cite{PhysRevA.87.061602}. 
A through analysis of the ground state phase diagram for soft-core interactions as in Eq.(\ref{MC}) 
was provided in Ref.~\cite{cinti_nat_comm} both for the high and low density limit using QMC techniques,
which are reviewed in Section~\ref{sec:2.2}.

In the high-density limit (when $n R_c^2 \gtrsim 1.8$) for low interaction strengths ($\alpha < 28$) 
the ground state of the system is in a uniform superfluid phase. 
Upon increasing the interaction at $\alpha \approx 28$ 
one crosses a first-order phase transition to a cluster supersolid phase characterized by a finite superfluid 
fraction and broken translational invariance where 
particles arrange in clusters (each cluster contains an average number of particles according to the density) 
in a triangular geometry. 
For even larger interactions $\alpha>38$ the ground state preserves 
triangular symmetry but superfluidity vanishes resulting into an uncorrelated cluster crystal.

In the intermediate density regime ($0.5 \lesssim n R_c^2 \lesssim 1.8$) one observes a direct superfuid to crystal
transition at commensurate lattice occupations where superfluidity vanishes abruptly. 
Simultaneously for incommensurate occupations of the crystalline lattice and mean field interactions in 
the range $28<\alpha<38$, superfluidity increases linearly with the fraction of vacancies  
or defects \cite{cinti_nat_comm} in nice agreement with the ALC scenario of defect delocalization.

In the very low density limit ($n R_c^2 \lesssim 0.5$) the physics of the system becomes largely irrelevant of the
detailed form of the soft core at small distances since the interparticle distance
is much higher than the range of the potential. Indeed, upon increasing the interaction at fixed density 
the system displays a first order phase transition from a superfluid to a crystalline phase as expected for a standard Van der Waals interaction. 

The elementary excitations in the mean field approximation are found by expanding the GP energy functional around
the solution $\psi_0({\bf r})$, obtaining the so called Bogoliubov de Gennes equations.
Denoting the change in $\psi(\mathbf{r},t)$ by $
\delta \psi(\mathbf{r},t)=e^{-i\mu t}\left[ u(\mathbf{r})
e^{-i\omega t}-v^*(\mathbf{r})e^{i\omega t}\right]$ 
and substituting this expression into the GPE Eq.(\ref{GPNH})
we find a set of two coupled linear differential equations:
\begin{equation}  \label{BdGnuni} 
\left\{
\begin{array}{ccl}
\left( -\frac{\nabla^2}{2} -\mu - \omega \right)u(\mathbf{r}) +
\alpha \int d\mathbf{r'} U(\mathbf{r}-\mathbf{r'})\left[\psi_0(\mathbf{r'})^2u(\mathbf{r})+
\psi_0(\mathbf{r})\psi_0(\mathbf{r'}) \left(u(\mathbf{r'})-v(\mathbf{r'})\right) \right] &=&0\\ \\
\left( -\frac{\nabla^2}{2} -\mu + \omega \right)v(\mathbf{r}) +
\alpha \int d\mathbf{r'} U(\mathbf{r}-\mathbf{r'})\left[\psi_0(\mathbf{r'})^2v(\mathbf{r})-
\psi_0(\mathbf{r})\psi_0(\mathbf{r'}) \left(u(\mathbf{r'})-v(\mathbf{r'})\right) \right] &=&0
\end{array} \right.
\end{equation}
for the Bogoliubov amplitudes $u(\mathbf{r})$ and $v(\mathbf{r})$.
The solution of Bogoliubov Eqs.(\ref{BdGnuni}) in the uniform superfluid phase is analytical:
\begin{equation} \label{BdGuni}
\epsilon_q =\sqrt{\frac{q^2}{2}\left(\frac{q^2}{2}+2\alpha\, \tilde U_q\right)},
\end{equation}
and depends only on the modulus of the excitation vector $\mathbf{q}$. Here $\tilde U_q$ is the Fourier
transform of the potential (see Eq.(\ref{ftrans}) for an expression in terms of  special functions).
The spectrum is linear for small momenta and the slope defines the sound 
velocity of the system; for $\alpha\ge 5.4$ one recovers the usual roton-maxon spectrum that is common
to other physical systems with non-local interactions as ultracold dipolar systems or superfluid $^4$He. 
In nonuniform phases one has to rely on a numerical solution of Eq.(\ref{BdGnuni}). Ref.\cite{PhysRevB.86.060510}
uses for example a Fourier expansion of the Bogoliubov amplitudes followed by a 
diagonalization of the corresponding equations.
The results that we present in Sec.(\ref{sec:3.2}) are instead obtained using a grid based 
solution in real space of  the Eqs.(\ref{BdGnuni}) 
for the lowest excitation bands and for $\mathbf{q}$ vectors
lying in the first Brillouin zone (FBZ) \cite{PhysRevA.87.061602} .

\subsection{Monte Carlo approach}
\label{sec:2.2} 

In order to assess the validity of mean field theory and extend these results to the regime
of lower densities we performed QMC calculations \cite{Ceperley1995} at finite temperature 
based on the worm algorithm \cite{Boninsegni2006L,Boninsegni2006}
in the canonical ensemble, carefully extrapolating  the zero temperature limit. 
We do not enter here into a detailed description of the algorithm to measure 
thermodynamic observables (the reader can refer to Ref.~\cite{Boninsegni2006} for an extensive summary of 
such techniques).
Here we focus on the application of QMC to recover information about the spectrum of the elementary
excitations of the system under study. 
In particular, it is possible to sample directly the imaginary-time intermediate scattering function
\begin{eqnarray}\label{Fqt}
F(\mathbf{k},\tau) =\langle{\hat\rho}_{\mathbf{k}}(\tau) {\hat\rho}^\dagger_{\mathbf{k}}(0)\rangle/N,
\end{eqnarray}
where the brackets denote a thermal average and
\begin{eqnarray}\label{Rho_K}
\hat\rho_\mathbf{k} = \sum_j e^{i \mathbf{k}  \cdot \mathbf{q}_j}
\end{eqnarray}
is the density fluctuation operator at wavevector $\mathbf{k}$.
The dynamic structure factor $S(\mathbf{k},\omega)$, which contains the information on the spectrum
of the elementary excitations of the 
density fluctuations, is related to $F(\mathbf{k},\tau)$ 
via an inverse Laplace transform:

\begin{eqnarray}\label{sqomega}
F(\mathbf{k},\tau) =\int \mathrm{d}\omega\, e^{-\tau\omega} S(\mathbf{k},\omega).
\end{eqnarray}
Here we face the well known ill-defined problem of inverting the Laplace transform from noisy data. 
There exists no general scheme to recover the exact inversion, however
some techniques manage to identify weight and frequency of the 
dominant contributions of a spectrum composed of well-defined peaks \cite{Vitali2010,Jarrell:1996fk}. 
However, previous investigations on bosonic systems with soft-core bosons with a step 
interaction potential \cite{PhysRevLett.108.175301} showed that fitting the $F(\mathbf{k},\tau)$ data 
directly via an $n$-pole approximation, i.e. assuming that the spectrum $S(\mathbf{k},\omega)$ 
is formed by a sum of $n$ delta functions of $\omega$ gives equally reliable results in good
quantitative agreement with more involved techniques based on the genetic inversion 
via falsification of the theories (GIFT) approach \cite{Vitali2010}. For such reasons in this work we 
focus on the $n$-pole approximation results to extract the excitation spectrum.

The estimator defined in Eq.(\ref{Fqt}) within the FBZ only contains information about longitudinal density fluctuations.
A study of the longitudinal modes was done for a class of soft-core potentials in \cite{PhysRevLett.108.175301}. 
The extension to the transverse excitations is not straightforward as it requires a careful analysis of the 
contributions to the intermediate
scattering function outside the FBZ. However, the spectral weight outside the FBZ is generically 
distributed on a greater number of modes, making difficult to perform the analytic continuation 
of the Laplace transform, as excitations near in energy tend to merge in the reconstructed spectra. 

In order to obtain the dispersion relation for transverse modes, one may measure imaginary-time current 
correlations \cite{PhysRevB.36.8343} and then perform the inverse Laplace transform. 
However, these current correlations estimators generically display much larger statistical errors than simple 
density correlations, making the analytical continuation unfeasible in most cases. 

Although it is possible to derive such current estimators with a reduced
statistical error within a method described in \cite{PhysRevLett.111.050406}, we use here a different and 
numerically less demanding strategy to couple 
transverse excitations in a lattice system to density fluctuations inside the FBZ.
The idea behind our method is to alter the imaginary time particle positions while applying the density
fluctuation operator so that transverse lattice displacements mimic longitudinal displacements, and vice-versa. 
For each slice in imaginary time we first identify the positions of the lattice sites of a triangular lattice to which particles belong (the nearest one). 
We then apply a weak potential to avoid the slow translations and rotations of the lattice as a whole 
that may occur in the simulations, so that their 
positions remain fixed. This potential is two orders of magnitude weaker than the average one between two 
lattice sites, so it does not alter the dispersion relations significantly.
For each time slice and particle position, we identify the nearest lattice site and rotate the particle position by 
${\pi}/{2}$  with respect to it. With this modified configuration we compute the density fluctuation operator defined in Eq.(\ref{Rho_K}) and from then on we follow the 
same procedure as in the standard case for obtaining excitation energies via Eq.(\ref{Fqt}) and the $n$-pole approximation.

The above operation transforms transverse lattice displacements so that they resemble longitudinal 
excitations.  
To show this, we first write the $j$ particle position as $\mathbf{q}_j=\mathbf{R}_j+\delta \mathbf{q}_j$, where 
$\mathbf{R}_j$ is the position of the closest lattice site to the particle labeled by index $j$ and $\delta\mathbf{q}_j$ 
is the relative displacement vector to the lattice site $\mathbf{R}_j$. 
Then we perform a local rotation of the relative displacement $\delta \mathbf{q}_j$ by ${\pi}/{2}$ with respect
to the corresponding lattice site $\mathbf{R}_j$ applying a rotation matrix:
\begin{equation}
M_{\pi/2} =
\left(
\begin{array}{cc}
0 & -1\\
1 & 0
\end{array} \right),
\end{equation}
and then calculate again the density operator:
\begin{equation} \label{mat}
\hat\rho_\mathbf{k}' = \sum_j e^{i \mathbf{k}  \cdot (\mathbf{R}_j+(M_{\pi/2}\, \delta\mathbf{q}_j))}= \sum_j e^{i (M_{\pi/2}^T\, \mathbf{k})  \cdot ((M_{\pi/2}^T\,\mathbf{R}_j)+\delta\mathbf{q}_j)},
\end{equation}
where the last term follows from the properties of the scalar product. 
Because of the presence of a rotation matrix, the term $M_{\pi/2}^T\, \mathbf{k} \cdot \delta\mathbf{q}_j$ is proportional to the component of $\delta\mathbf{q}_j$ orthogonal to $\mathbf{k}$, while of course in the absence of any rotation it would be proportional the the component of $\delta\mathbf{q}_j$ parallel to $\mathbf{k}$. Transverse excitations are therefore recorded in the transformed density fluctuation operator $\hat\rho_\mathbf{k}'$ even if $\mathbf{k}$ lies within the FBZ.

As expected, the present reasoning is based on the assumption that the crystalline structure exists and 
that we can unambiguously identify for each particle the lattice site to which it is pertaining. 
It is worth stressing that the configuration produced by the transformation defined above is 
merely a trick used to couple a term like the one in Eq.(\ref{Rho_K}) to transverse excitations and leaves
unaffected the internal Monte Carlo dynamics.

\begin{figure} 
\begin{center}
\includegraphics[height=.4\textheight]{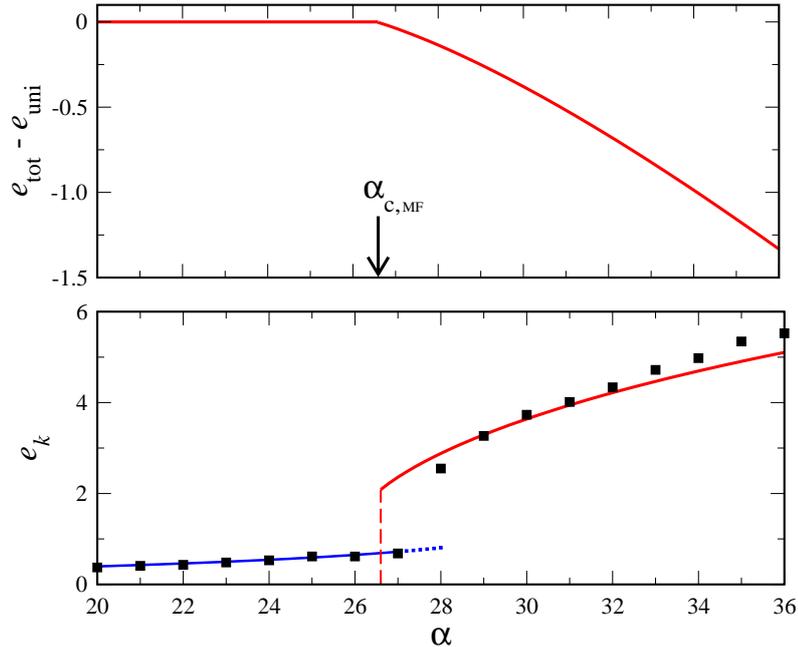}
\end{center}
\caption{\label{energy} Upper panel: Energy difference between the uniform state and 
the modulated state in the mean field approach obtained by 
minimizing the Gross-Pitaevskii energy functional in Eq.(\ref{GPenergy}). 
The SF-SS transition is located at $\alpha=26.6$ for the value of
lattice spacing equal to $a/R_c=1.6$. 
Lower panel: kinetic energy per particle from QMC calculations for 400 
particles at density $n R_c^2=4.5$. The red line is the kinetic term of 
Eq.(\ref{GPenergy}). In this approximation in the uniform phase kinetic energy vanishes. 
The blue line is a beyond mean field calculation of the kinetic energy based on Eq.(\ref{kin_bog}) 
for density $n\, R_c^2=4.5$ that well agrees with QMC data. Error bars on QMC data are smaller
than the dimension of the points.}
\end{figure}

\section{Results}
\label{sec:3}

\subsection{Ground states properties}
\label{sec:3.1} 

In the upper panel of Fig.~\ref{energy} we report the mean field energy 
per particle of the uniform phase $e_{uni}=2\pi^2 / 3 \sqrt{3}$ otained with full 
numerical minimization of Eq.(\ref{GPenergy}).
For $\alpha<\alpha_{c,MF}=26.6$ the density is homogeneous and the system is then superfluid; 
upon increasing the interaction one crosses a first order phase transition to a phase
with a modulated density. In the lower panel  (again Fig.~\ref{energy}) 
we show the kinetic energy per particle comparing QMC results extrapolated to the $T\to0$ limit,
and the values from mean field. 
QMC data clearly display an abrupt jump of the kinetic energy for $27<\alpha<28$ signaling the transition to
the supersolid phase in good agreement
with the mean field results within few percents.

We also included Bogoliubov fluctuations energy on top of the uniform solution  
that is displayed as the solid line from 
$\alpha=20$ to $\alpha\approx27$ in Fig.~\ref{energy} (lower panel). This calculation is 
based on the evaluation of the following integral \cite{pethick}:
\begin{equation} \label{kin_bog}
e_{kin} = \frac{1}{8\pi n} \int\limits_{0}^\infty dq\; q^3 \left(\frac{\frac{q^2}{2}+\alpha\, \tilde U_q}{\sqrt{\frac{q^2}{2}\left(\frac{q^2}{2}+2\alpha\, \tilde U_q\right)}} 
- 1\right),
\end{equation}
where $\tilde U_q$ is the Fourier transform of the interaction potential:
\begin{equation} \label{ftrans}
\tilde U_q = \frac{\pi}{3} \mathbf{G}_{0,6}^{4,0} \left(
\begin{array}{c|cccccc}
\frac{q^6}{46656}&0&\frac{1}{3}&\frac{2}{3}&\frac{2}{3}&0&\frac{1}{3}
\end{array}
\right)
\end{equation} 
and $\mathbf{G}_{p,q}^{m,n}\left(z\Big|
\begin{array}{c}
a_1,\dots,a_p\\
b_1,\dots, b_q
\end{array}
\right)$ is the Meijer's $G$-function \cite{gradshteyn2007}.

 
We see that Eq.(\ref{kin_bog}) matches very well the QMC calculations performed in the superfluid phase.
These results can be understood noting that the condensate fraction is relatively high for this class of
soft-core potentials when the density is sufficiently high 
\cite{PhysRevA.87.061602,PhysRevA.3.1067,PhysRevA.71.023605}. 
In order to verify this assertion, we calculated the depletion $n_{exc}/n$  
in the superfluid phase within the Bogoliubov approximation and obtained the largest value 
$n_{exc}/n \approx 38\%$ close to the transition.


%
%

\subsection{Excitations}
\label{sec:3.2} 

\begin{figure} 
\begin{center}
\includegraphics[height=.28\textheight]{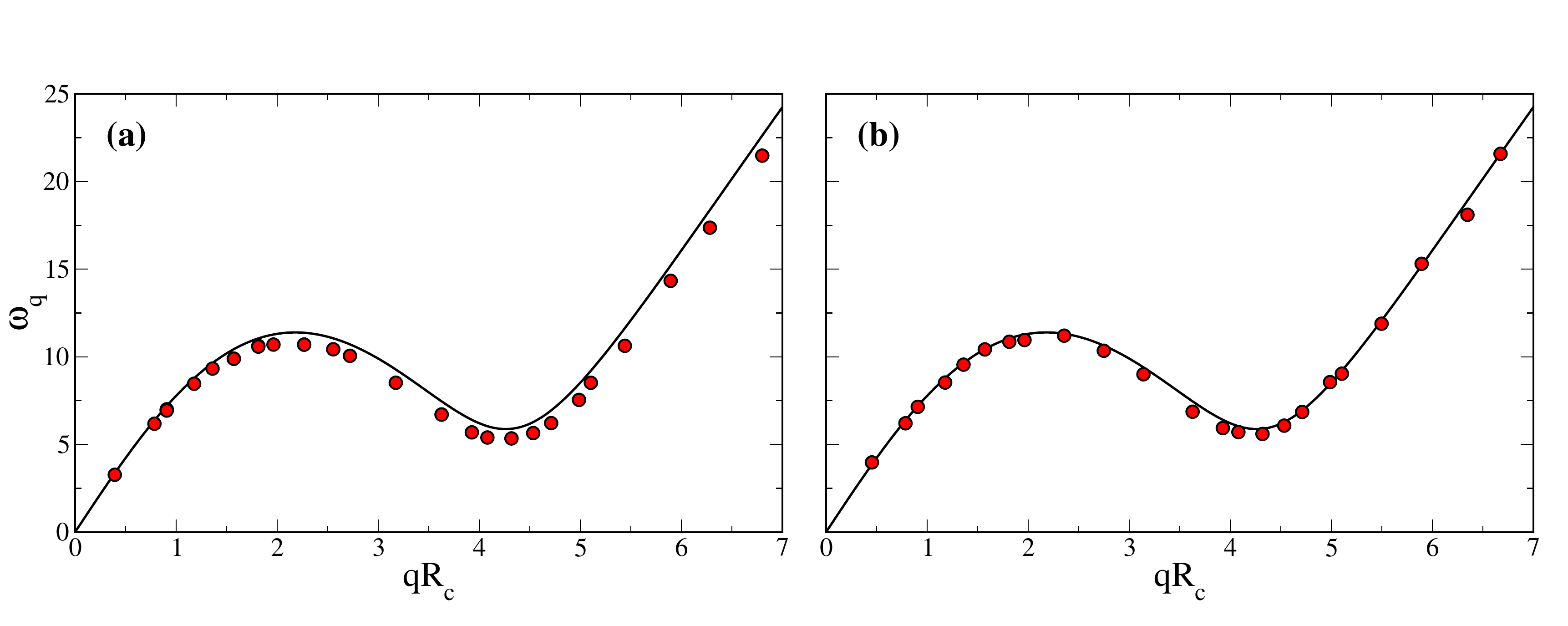}
\end{center}
\caption{\label{excitations1}Spectra in the superfluid phase at $\alpha=20$ with the typical roton-maxon feature.  
Dots correspond to the one-pole approximations 
and are compared with the solution of Bogoliubov-de Gennes Eqs.(\ref{BdGuni}) (full line). 
Excitation energy is in units of $\hbar^2/m R_c^2$.
\textbf{(a)} Density $n\, R_c^2=1.12$. \textbf{(b)} $n\, R_c^2=3.8$.
Error bars on the frequencies are recovered from the one-pole approximation and are smaller than the point size.}
\end{figure}

In Fig.~\ref{excitations1} we show the dispersion relations, $\omega_{q}$ versus $qR_c$, obtained applying 
the one-pole approximation (points, within QMC simulations, see Section~\ref{sec:2.2}),
and the analytical mean-field solution of Bogoliubov (see Eqs.(\ref{BdGuni}), continuous lines)
at low (Fig.~\ref{excitations1}a, $n\, R_c^2=1.12$) and high density (Fig.~\ref{excitations1}b, $n\, R_c^2=3.8$),
respectively. Both panels present results at the same mean field interaction strength, $\alpha=20$. Such an interaction
turns out to be high enough for exhibiting an unambiguous roton instability. 
In Fig.~\ref{excitations1}a the simulations have been performed using 250 particles
while at the higher density (Fig.~\ref{excitations1}b) we have used 850 particles.
Mean-field calculations quantitatively agree with QMC simulations
with a small deviation (about 5$\%$) between the two approaches for the lowest density at higher momenta. 
We interpret this disagreement as an effect of the stronger correlations in the low density limit \cite{cinti_nat_comm}.

\begin{figure}[b!]
\begin{center}
\includegraphics[height=.3\textheight]{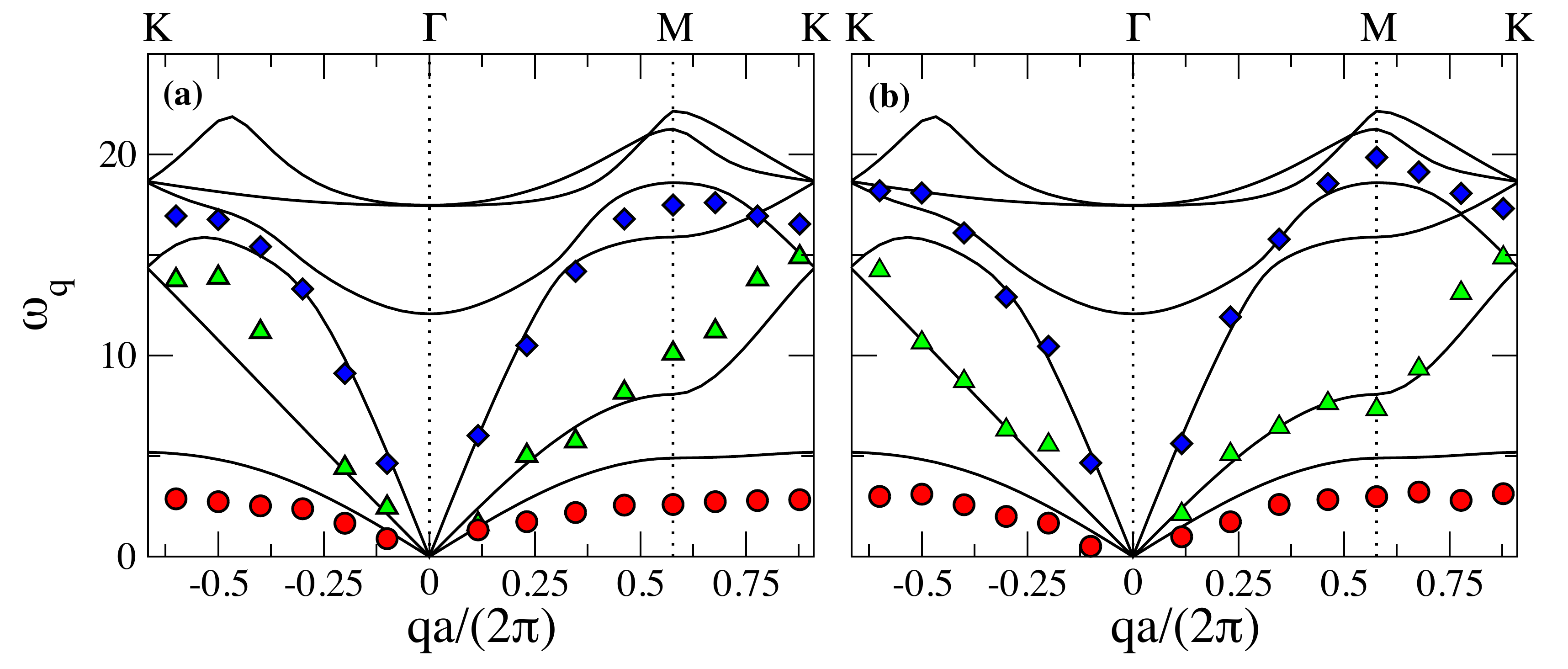}
\end{center}
\caption{\label{exSS}Excitation spectra in the supersolid phase with an interaction 
strength $\alpha=34$. 
Dots correspond to the $n$-pole approximations,
red circle Bogoliubov band, blue diamond longitudinal modes, and green triangle transverse modes.
They are compared with the numerical solution of Bogoliubov-de Gennes Eqs.(\ref{BdGnuni}) (full lines). 
Excitation energy is in units of $\hbar^2/m R_c^2$.
\textbf{(a)} Density $nR_c^2=1.12$.
\textbf{(b)} Density $nR_c^2=3.8$.}
\end{figure}

In Fig.~\ref{exSS}, we plot the excitation spectra in the supersolid phase at mean field interaction
$\alpha=34$ with the same densities as in Fig.~\ref{excitations1}. 
Mean-field theory predicts three Goldstone modes that reflect the three spontaneously broken symmetries 
in a supersolid phase. 
The lowest band corresponds to the Bogoliubov mode that appears due to the off-diagonal 
long range order of the one body density matrix.
The other two gapless bands are a consequence of the broken translational symmetry of the crystalline structure.
Specifically, the highest gapless mode is a longitudinal band, whereas the intermediate mode is a transverse excitation
\cite{PhysRevA.87.061602}.

\begin{figure}[t!]
\begin{center}
\includegraphics[height=.3\textheight]{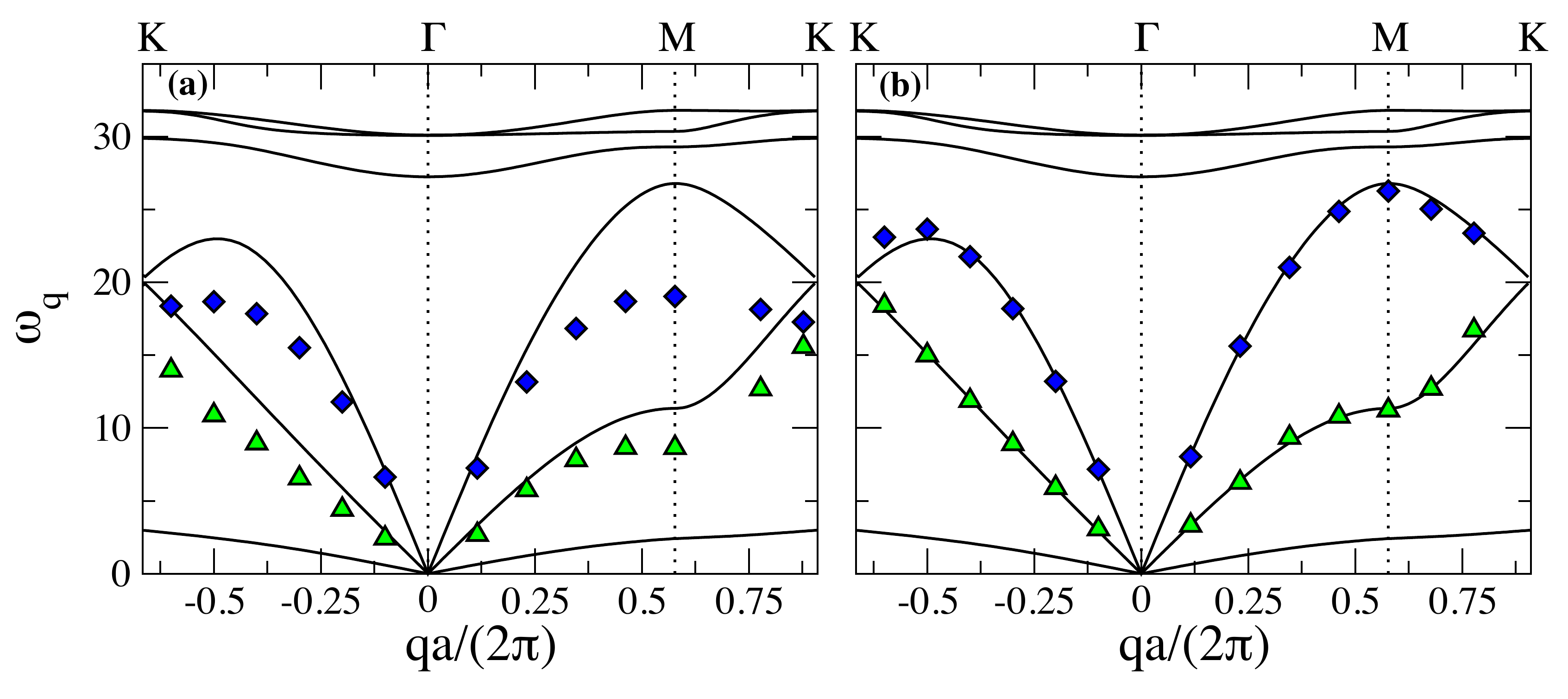}
\end{center}
\caption{\label{exCR}
Excitation spectra in the crystal phase with an interaction 
strength $\alpha=60$. 
Dots correspond to the $n$-pole approximations,
blue diamond longitudinal modes, and green triangle transverse modes.
They are compared with the numerical solution of Bogoliubov-de Gennes Eqs.(\ref{BdGnuni}) (full lines). 
Excitation energy is in units of $\hbar^2/m R_c^2$.
\textbf{(a)} Density $nR_c^2=1.12$.
\textbf{(b)} Density $nR_c^2=3.8$.}
\end{figure}

At high filling (Fig.~\ref{exSS}b) one can notice a satisfactory agreement between mean-field results and 
QMC spectra, in particular for the transverse band. 
Nevertheless, due to its limited resolution at higher energies, QMC 
does not allow to distinguish among longitudinal mode contributions
and  optical bands, since they are very close in energy for intermediate interactions. 
It is worth mentioning that, for both densities, mean-field theory predicts a larger Bogoliubov branch
compared to QMC points.
We mention that we have observed a similar behaviour in the entire supersolid regime.
Finally, at lower densities the mean field theory still qualitatively agrees with QMC, 
even though the transverse band deviates systematically to larger energies if confronted to higher densities.
 
Now we discuss the excitation spectrum in the crystalline phase. QMC shows that for 
interactions $\alpha \gtrsim 38$ superfluidity vanishes as well as the
contribution of the Bogoliubov band to the structure factor. Fig.~\ref{exCR}a and Fig.~\ref{exCR}b report
the energy of the elementary excitations at $\alpha=60$, again for the same densities as in Fig.~\ref{excitations1}. 
We observe that the Bogoliubov mode is still present from mean field calculations 
due to the global coherency assumed in the  
derivation of Eqs.(\ref{BdGnuni}) even for very large interactions.
Simultaneously at large densities gapless modes are far in energy from the optical
branches. We indeed see that Fig.~\ref{exCR}b provides a good matching 
between the prediction of longitudinal and transverse modes by QMC and the approach 
of Section~\ref{sec:2.1}. Finally, we observe that in Fig.~\ref{exCR}a correlations limit the predictive 
power of the Bogoliubov equations that show deviations up to $\approx 30\%$ for large momenta.

\section{Conclusion}
\label{sec:4} 

In the present paper we analyzed in detail some significant properties of soft-core 
bosons interacting through a soft-core potential that can be realized with Rydberg-dressed atoms. 
In particular, we considered a two dimensional system at zero temperature,
with an interaction range such that the ground state displays a supersolid phase.
In particular, we extended some previous studies \cite{PhysRevLett.108.175301,PhysRevA.87.061602} 
by calculating the transverse gapless mode within a QMC approach.
This technique allows to evaluate accurately the transverse excitations in a 
translationally broken phase, with statistical errors comparable 
to the ones obtained by the standard procedure used for longitudinal excitations \cite{PhysRevLett.108.175301}.
  
In the regime of density-modulated superfluid, we have compared mean-field results with
exact QMC calculations, verifying that the ground state and the elementary 
excitations are quantitatively described by a non-local Gross-Pitaevskii equation 
supplemented by the fluctuations encoded in the Bogoliubov formalism. 
Within this framework, we have included the Bogoliubov fluctuations and checked that 
the average kinetic energy is fully consistent with 
the QMC calculations in the high density limit. 
Yet, a comparison between the numerical simulation based on the 
calculation of the intermediate scattering function
and the solutions of the Bogoliubov-de Gennes 
equations in the supersolid phase reveals a convincing agreement, as previously observed for 
the longitudinal phonon and Bogoliubov modes in the case of a step-like interaction potential \cite{PhysRevA.87.061602}.
On the other hand the global phase coherence assumed in the mean field approach overestimates the energy of the 
Bogoliubov excitation band with respect to QMC calculations, both in the supersolid  
and in the crystalline phase and it can be as large as $30\%$ for large momenta. 

Concerning the low density regime, i.e. where quantum fluctuations are expected to play a major role, we conclude that
mean field computations still quantitatively predict the spectra for a uniform superfluid for a wide range of momenta
and interactions. At the same time, for supersolid and crystalline phases the deviations from QMC become larger 
upon increasing the interactions, reaching approximately
 $30\%$ for an interaction strength $\alpha=60$ and density $n R_c^2=1.12$.

The predictions made in this work may serve as a guidance for further theoretical and experimental studies
on supersolidity with laser excited Bose-Einstein condensates. A natural extension of this work, for instance, 
might regard the behaviour of soft-core bosons in three dimensions in the low density regime to complement and extend 
recent analysis on cluster supersolidity \cite{Henkel10,PhysRevLett.106.170401,PhysRevA.88.033618} 
which would though require more refined techniques to calculate transverse modes in inhomogeneous phases. 
Finally we mention that the analysis of the spectrum could provide a natural 
way to observe supersolidity in the laboratory \cite{PhysRevLett.83.2876,PhysRevLett.88.120407,FixMe}. Several 
experimental techniques have indeed been recently tested and implemented 
in ultra-cold atomic systems to detect elementary excitations, including Bragg \cite{PhysRevLett.101.135301,PhysRevLett.102.155301} and 
Raman spectroscopy \cite{PhysRevLett.98.240402}.



\begin{acknowledgements}
We acknowledge M. Boninsegni, G. Carleo, S. Moroni, and T. Pohl for valuable discussions.
\end{acknowledgements}

\bibliographystyle{spphys}       
\bibliography{./bose}   

%
%


\end{document}